\documentclass[journal,transmag]{IEEEtran}
\IEEEoverridecommandlockouts
\usepackage{cite}
\usepackage{amsmath,amssymb,amsfonts}
\usepackage{algorithmic}
\usepackage{graphicx}
\usepackage{textcomp}
\usepackage{xcolor}

\usepackage{url}

\usepackage{breakurl}
\usepackage[breaklinks]{hyperref}

\usepackage{array}
\usepackage{booktabs}
\usepackage{mathtools, cuted}
\usepackage{multirow}
\newcommand{\ignore}[2]{\hspace{0in}#2}
\usepackage{subcaption}
\usepackage{algorithm}
\usepackage{algorithmic}

\def\BibTeX{{\rm B\kern-.05em{\sc i\kern-.025em b}\kern-.08em
    T\kern-.1667em\lower.7ex\hbox{E}\kern-.125emX}}
\begin{document}

\title{Fuzzy Logic-based Robust Failure Handling Mechanism for Fog Computing
}

\author{\IEEEauthorblockN{Ranesh Naha, Saurabh Garg, Sudheer Kumar Battula, Muhammad Bilal Amin}
\IEEEauthorblockA{\textit{School of Technology, Environments and Design} \\
\textit{University of Tasmania}\\
Hobart, Australia \\
\{raneshkumar.naha,saurabh.garg,sudheerkumar.battula,bilal.amin\}@utas.edu.au}
\and
\IEEEauthorblockN{Rajiv Ranjan}
\IEEEauthorblockA{\textit{School of Computing} \\
\textit{Newcastle University}\\
Newcastle, United Kingdom \\
raj.ranjan@ncl.ac.uk}

}

\maketitle

\begin{abstract}
\ignore{**Introduction**} Fog computing is an emerging computing paradigm which is mostly suitable for time-sensitive and real-time Internet of Things (IoT) applications. Academia and industries are focusing on the exploration of various aspects of Fog computing for market adoption. The key idea of the Fog computing paradigm is to use idle computation resources of various handheld, mobile, stationary and network devices around us to serve the application requests in the Fog-IoT environment. \ignore{**Problem Statement**}The devices in the Fog environment are autonomous and not exclusively dedicated for Fog application processing, with high the probability of failure. To handle failure, there are several techniques available in the literature such as checkpointing, task migration; each of which work well in different scenarios and does not guarantee a robust schedule. In contrast, this work focused on solving the problem of managing application failure in the Fog environment by proposing a composite solution (combining fuzzy logic-based task checkpointing and task migration techniques with task replication) for failure handling and generating a robust schedule. \ignore{**evaluation/Results/Conclusion**} We evaluate the proposed methods using real failure traces in terms of application execution time, delay and cost. Average delay and total processing time improved by 56\% and 48\%  respectively on an average for the proposed solution, compared with the existing failure handling approaches. \ignore{??EFFECTIVENESS??}\ignore{availability and SLA violation rate}
\end{abstract}

%


\begin{IEEEkeywords}
Fog Computing, Application Failure, Dynamic Resource, Fault Tolerance, Robustness
\end{IEEEkeywords}

\section{Introduction}
Fog is a highly distributed environment in which numerous autonomous devices are contributing to process application requests which are known as Fog device \cite{naha2018Fog}. The contributing devices can be benefited financially by allowing Fog platform to use their resources for Fog application processing \cite{battula2019micro}. The devices in this environment are not completely dedicated to Fog application processing \cite{yi2015survey}. 
Hence, there is no guarantee that devices are always available for application processing. The device can even fail after starting the processing of the Fog application. In such a scenario, it is important to make application scheduling robust for successful application execution in the Fog environment, despite any failure of Fog devices during the execution of the application. This will reduce the impact of application failure in the Fog environment. 


In order to meet the time sensitivity of the applications, handling failure in the Fog environment is an important and challenging task \cite{bermbach2017research}. For application processing, the cloud computing environment mostly depends on the cloud data centre \cite{buyya2009cloud} where the rate of failure is not that high compared with the Fog environment. Fog devices are controlled by decentralised entities while cloud data centres are managed by some central entities. Hence, predicting application failure in the cloud environment is less complicated compared with such predictions in Fog computing. In the Fog environment, it is difficult to predict the failure of the computation resources due to the unstable characteristics of the available resources in the Fog devices. Hence, a robust scheduling algorithm needs to be developed. On the other hand, to prove the correctness of the robust scheduling, an evaluation with real failure traces is required.

Two methods are available for handling failures in service-oriented computing; these are proactive and reactive failure handling mechanisms  \cite{gill2018failure} \cite{sharma2016reliability}. Proactive failure needs to be considered in a highly distributed environment in such a way that failure handling has taken place before occurrence of the failure. However, this solution will not always be suitable because some failures might occur beyond our prediction. This is because of the possibilities of device malfunction, user interruption and uncertain changes in resource availability. Therefore, we need to consider reactive failure handling which usually takes place after the failure has occurred. Due to the unstable nature of Fog devices, applying either of the methods is not always useful for ensuring the successful execution of the application in a time-sensitive manner. Hence, we are using task replication methods along with proactive and reactive failure handling methods.


In this paper, we propose a composite solution by using proactive and reactive failure handling methods with replication. The key contributions of this paper are as follows:

\begin{enumerate}
    \item We propose a fuzzy logic-based method to deal with unpredicted and predicted failures.
    \item Our failure handling method consider time sensitivity of the application, as well as dynamic changes in the available resources in the devices.
    \item We evaluate the proposed failure handling method using real failure traces.
\end{enumerate}

 The rest of the paper is organized as follows: Section II presents related literature of failure handling mechanisms in P2P system, cluster, grid, cloud and Fog computing paradigm. Section III discusses the application scenario for the proposed solution. Section IV discusses the definition of the problem. A brief description of various resources failures and their solutions presents in Section V. Section VI explains a detailed description of proposed Fuzzy logic-based solution. Section VII discusses the experimental setup and evaluation technique. Section VIII presents experimental results with discussion. Finally, section IX concludes the paper.

\section{Related Work}
This section describes some related work on failure handling mechanisms in different distributed computing paradigms. We have chosen to survey failure handling methods in order to verify the uniqueness of our research. We reviewed some methods that are being used for the P2P system, as well as cluster, grid, cloud and Fog computing paradigms, in order to make the system fault-tolerant.

\subsection{Failure handling in P2P System}
Samant and Bhattacharyya \cite{samant2004topology} examine the impact of node failure on the accessibility of resources on P2P networks. Their work examines how search efforts, topology of the network and redundant resource can influence accessibility when various level node failure takes place. Vanthournout and Deconinck \cite{vanthournout2004building} proposed three strategies to realise \ignore{this: the use of a small-world topology,} the use of self- organisation mechanisms for failure handling and failure detection. \ignore{, and the use of cross-partition pointers to deal with network partitions}.

Lin et al. \cite{lin2007fault} presented an efficient fault-tolerant approach for the super-peers of peer-to-peer (P2P) file-sharing systems. \ignore{In the super-peer based P2P file-sharing system, peers are organised into multiple groups. Each group has a special peer called a super-peer to serve the regular peers within the group. In this hierarchical architecture, if the super-peer departs (fails), any file queries to its regular serving peers cannot be delivered. In their proposed approach, a multiple publication technique to make each regular peer logically connect with two or more super peers in other groups is suggested.} Mawlood-Yunis et al. \cite{mawlood2010p2p} identify the disconnection failure problem, due to temporary semantic mapping faults, and propose an game theory based Fault Tolerant Adaptive Query Routing (FTAQR) algorithm to resolve it. \ignore{To identify the problem, the work used a simulation model of SP2P systems. The Fault Tolerant Adaptive Query Routing (FTAQR) algorithm, which was proposed to resolve the problem, is an adaptation of the generous tit-for-tat method originally developed in evolutionary game theory.}


\subsection{Failure handling in Cluster}
Li and Lan \cite{li2006exploit} proposed FT-Pro, an adaptive fault management mechanism that optimally chooses migration, checkpointing or no action to reduce the application execution time in the presence of failures based on the failure prediction. \ignore{A cost-based evaluation model is presented for dynamic decisions at run-time. Using the actual failure log from a production cluster at NCSA, Li and Lan demonstrate that, even with modest failure prediction accuracy.} Various methods have been used in cluster computing to predict failure event. These methods include genetic algorithms  \cite{weiss1998learning}, rule-based data mining \cite{vilalta2002predicting1, sahoo2003critical, vilalta2002predicting} and Markov chain \cite{hoffmann2011advanced}. Many other works focus on fault management techniques which are based on prediction. Leangsuksun et al. \cite{leangsuksun2004failure} proposed a predictive failure handling mechanism which scheduled deliberate job checkpointing and migration. 
In another work, Castelli et al. \cite{castelli2001proactive} employ a different approach to failure prediction. In their approach, 
they first predict the resource exhaustion failure proactively and then conduct software rejuvenation. To maximise system throughput, Oliner et al. \cite{oliner2005probabilistic} used the coordinative checkpointing strategy that optimistically skips a checkpoint when no failure prediction exists in the near future. Chakravorty et al. \cite{chakravorty2005proactive} proposed software-based prediction of failure which basically migrates task before the failure actually occurs.

\subsection{Failure handling in Grid}
Hwang and Kesselman \cite{hwang2003grid} proposed a flexible failure handling framework for the grid which is comprised of two phases: failure detection and recovery phases. In the failure detection phase, an event notification mechanism notifies failures. A failure handler deals with the failures at two levels: the task level and the workflow level. Task level failures are handled by retrying, checkpointing and replication. At the workflow level, they are managed by alternative task and redundancy. 
Jin et al. \cite{jin2003fault} proposed the Fault Tolerant Grid Platform (FTGP) approach from the perspective of grid users, taking the nature of grid faults into account. \ignore{Based on Globus fault detection services, a high-level service is proposed to support fault-handling strategies for all the grid components, including grid nodes, system-level and application-level software components. However, the work only considers atomic tasks and applications that are unrelated to each other, and concentrate on fault-tolerant strategies in computational grids.}

Lee et al. \cite{lee2005resource} proposed a resource manager for optimal resource selection. The proposed resource manager automatically selects a set of optimal resources from candidate resources which achieve optimal performance using a genetic algorithm with a fault tolerance service that satisfies QoS requirements. Lee et al. implemented a fault detector and a fault manager which will handle failure by job migration, using a checkpoint. Kandaswamy et al. \cite{kandaswamy2008fault} proposed a fault-tolerance technique using over-provisioning and migration for computational grids. \ignore{Over-provisioning or replication is a fault-tolerance mechanism in which multiple copies of an application (with the same input data-set) are executed in parallel. The idea is to maximise the probability of success for the application so that if one copy fails, then another copy may succeed. The proposed over-provisioning algorithm in this paper determines the best set of resources to replicate the application. In migration, an application is progressively restarted from the last good checkpoint (if available) on a different resource, in case of failures. The migration algorithm determines the best migration path.} Khoo and Veeravalli \cite{khoo2010pro} proposed a failure handling mechanism based on pro-active failure handling strategies for Grid environments. \ignore{These strategies estimate the availability of resources in the Grid, and also pre-emptively calculate the expected long term capacity of the Grid. Using these strategies, Khoo and Veeravalli created modified versions of the backfill and replication algorithms to include all three pro-active strategies, in order to ascertain the effectiveness of each of them in the prevention of job failures during execution. However, they have not considered the failures that occur during runtime.}

\subsection{Failure handling in cloud computing}
Much research on handling failures in the cloud environment has been undertaken to provide a failure-prone environment. Two review works \cite{gill2018failure,sharma2016reliability}, in which all kinds of failures are categorised into reactive and proactive failure methods, extensively evaluated failure handling mechanisms in the cloud. Reactive failure mechanisms were further divided into three sub-categories: checkpointing, replication and logging. While VM migration was considered as proactive failure management, Gill and Buyya \cite{gill2018failure} suggested continuous monitoring of resource allocation to manage failures in the cloud environment during operation. Battula et al. \cite{battula2019efficient} proposed an efficient resource monitoring service for Fog computing which suggesting failure handling is essential for an efficient resource management in Fog environment. Sharma et al. \cite{sharma2016reliability} point out that predicting resource behaviour is critical in the cloud environment.  

Sharma et al. \cite{sharma2019failure} proposed a failure-aware VM consolidation technique based on exponential smoothing. They employ checkpointing and migration of VM in their proposed method. Luo et al. \cite{luo2019improving} proposed a Markov-based model to examine failures in large-scale cloud computing systems. They employ reliability-aware resource scheduling to improve fault tolerance. Although cloud computing is a mature, it still has a lack of service reliability. Hence, Buyya et al. \cite{buyya2019manifesto} suggested investigating failure-aware provisioning and reliability-aware service management mechanisms for the cloud computing environment.

\subsection{Failure handling in Fog computing}
Existing failure handling methods in P2P, distributed and cloud computing mainly used extra resources to cover failure but in Fog computing, fault tolerance is challenging due to some unfavourable factors such as resource constraints and multiple procedures  \cite{liu2017framework}. Most of those methods considered only one failure handling method (for example, checkpointing or replication or resubmission) for fault tolerance \cite{alarifi2019fault}. Also, they did not consider any time sensitivity of the user request \cite{alarifi2019fault}. Hence, some researchers proposed new methods for failure handling in the Fog computing environment \cite{alarifi2019fault,tajiki2019software}.

A fault-tolerant scheduling method (FTSM) is proposed by Alarifi et al. \cite{alarifi2019fault} for the Fog-Cloud environment. In their approach, the system submits time-tolerant requests to the cloud and time-sensitive requests to the edge devices. FTSM finds the checkpoint interval based on the operation time between failures for the devices. However, Alarifi et al. \cite{alarifi2019fault} did not consider the prediction of the failure for devices based on the fluctuating availability of computation resources in the devices. Tajiki et al. \cite{tajiki2019software} proposed the Heuristic Fast Failure Recovery (HFFR) algorithm for software-defined service function chaining for Fog computing with failure consideration. The main idea of their proposed method is to find failure probability based on the predefined threshold. Similar to FTSM, HFFR did not consider the dynamic changes in the available resources. Also, both works do not consider real failure traces for evaluating their proposed methods.  

In summary, existing failure handling methods in Fog computing did not take into account fully the dynamic availability of Fog resources. In this paper, we propose a combined approach of proactive and reactive failure handling with task replication to tackle highly dynamic behavior of Fog resources. Hence, this research is carried out to propose a composite solution of utilising proactive, reactive and replication failure handling methods with dynamic changes of the resources in the Fog devices.





\section{Application Scenario}
To motivate the problem solved in this paper, let us assume that an emergency vehicle is using a smart transportation application and moving from point A to point B. The vehicle has to choose the shortest route to the destination.
To fulfill this requirement the system needs to process data generated or stored in a dash cam, surveillance camera and sensors. Based on the traffic conditions, the following actions need to be taken: (i) inform other vehicles ahead that an emergency vehicle is approaching; (ii) override signals if there are multiple road junctions along the way; (iii) do the relevant processing in the Fog devices, and (iv) take action following the processing.


Other incidents might also occur while the emergency vehicle is en route. The system should act promptly to minimise the delay in reaching the destination. Here, the system needs to process data from sensors as well as video data from dashcams and surveillance cameras. All of the processing for the above application scenarios is done in Fog devices to comply with the time sensitivity. Therefore, the utilisation of processing power and on-time processing is important. It is possible to ensure time sensitivity of the application by distributing the application workload among Fog devices \ignore{\cite{} mywork}. But the issue is what will happen if the Fog node has failed? We need to ensure that the outcome of the application should meet time-sensitive requirements in which the robustness of the scheduling approach will be assured. 

Robustness is a feature of the scheduling process in which application execution will be successful by ensuring time sensitivity, even if the resource has failed, any errors have occurred in the system components or any erroneous input has taken place. In our application scenario, the application always requests the completion of the processing by defining some deadline. However, our concern is how to deal with the failure of the resources during operation. We are specifically focused on minimising the impact of the failure on the applications, due to resource failure, by handling the situations in which Fog device resources have failed.




\section{Problem Definition}

This research is carried out to solve the following problem: How to meet user requirements for applications in the Fog environment, with a consideration of device failure, in order to satisfy any time-sensitive requirements of the application, while available resources in the devices are changing dynamically?



Scheduling all related tasks to Fog devices is not that complicated task if we can assume that all devices are up and running and there are no chances of failure of the Fog devices. But in reality, the chance of failure in the Fog environment is very high since the devices are not dedicated to running Fog applications. On the other hand, most of the devices in the Fog depend on wireless connectivity. Also, the devices are mobile and are moving from one cluster to another very frequently. Next, most of the Fog devices are not stationary, meaning that the devices have limited battery power. Furthermore, the application might be interrupted by the owner of the devices (for example, the owner turns off the device for some reason, the owner does not want to participate at that moment or the owner wants to run another application which requires some resources to be freed up). Due to all of the above reasons, the chances of failure of computation resources are very much higher than in any other distributed system. To ensure the robustness of the scheduling algorithm, we need to deal with resource failures in a way that the application user would not affected.



\section{Resource Failure and Counter-measures}
The resources might fail in the Fog environment for many reasons. The reasons for failure can be categorised, such as the termination of the application to run the native application, network failure, the device being moved to another cluster, power outage, human interruption, software and hardware failure, and network attacks. Due to the dynamicity of the available resources in the devices and their mobility, we can categorise all types of failure into two basic types: (i) unpredicted/immediate failure and (ii) predicted failure.



We can handle failures in two different ways. Firstly, we can manage the resource failure after it took place which is referred to as reactive failure. Secondly, it is possible to have countermeasures before the occurrence of the resource failure; this is known as proactive failure handling. Both types of failure handling mechanism have different approaches to manage resource failures.

In a reactive failure handling mechanism, we can employ checkpointing and replication. In application checkpointing the state of the application is stored in reliable storage and, if the application has failed, it does not need to rerun the application from the beginning. It will start the application from the point where the latest state has been saved. There are two types of checkpointing: i) coordinated or periodic checkpointing and ii) uncoordinated or random checkpointing. In coordinated checkpointing, the checkpoint should be consistent for the processes. In uncoordinated checkpointing, each process checkpoints its state. The other type of reactive failure handling mechanism is replication which always run replicas of the running processes in different devices.

The basic way to solve immediate failure is re-running the whole application but this is not a good way to solve the problem. For example, if a certain percentage of processing is done there is no point in processing the same portion of the application. Hence, the only solution for immediate failure is checkpointing. Some researchers have argued that checkpointing is not a good solution for the Fog environment because the Fog is a highly dynamic environment \cite{yi2015survey,kai2016Fog}. Yi et al. \cite{yi2015survey} suggested that replications are more suitable for the Fog but multiple Fog nodes need to be working together. Madsen et al. \cite{madsen2013reliability} suggested using checkpointing for the Fog which will save computation time for the faulty tasks. Some researchers used checkpointing in the Fog as a fault-tolerant technique  \cite{de2018fault,puliafito2018companion}. We need to ascertain if there is any way to accommodate checkpointing in the Fog environment. To do that we need to evaluate our solution in simulation and in a real environment as well. We evaluate our proposed method in a simulated environment with real failure traces. 

In a proactive failure handling process, we can employ the migration process before the resource failed. Since the Fog is for time-sensitive applications, we need to migrate the application without disconnecting devices. Hence, we need to employ live migration for this process. Two basic types of migration are i) pre-copy migration and ii) post-copy migration  \cite{sharma2016reliability}. In post-copy migration, application migration needs to be initiated by suspending the application at the source which will increase down-time. To minimise downtime, pre-copy migration needs to be employed. To resolve the predicted failure, we can employ pre-copy live migration. Once we can predict that an application is going to fail then we can migrate the application to another Fog device. But again the question is raised: how to decide when and where to migrate? \ignore{Thus, we need to answer the following two basic questions about application migration: when to migrate? where to migrate to?} However, this research only dealt with when to migrate not where to migrate to. In the Fog computing environment, the chances of both predicted and unpredicted failures are high because of the limited and highly dynamic resource availability in the devices. To ensure the robustness of the scheduling algorithm, we need to handle both predicted and unpredicted failures which will minimise their impact.

\section{Fuzzy logic-based failure handling mechanism}
To handle predicted and unpredicted failure we employ the fuzzy logic-based solution. Classical logic usually has a bivalent proposition, which may be true or false in principle. On the other hand, fuzzy logic can represent actual membership of both true and false for a function. Some propositions might be true and false to a certain degree, rather than being true or false only. For example, for a Fog device, mobility, response time and power availability might cause the failure of a device. However, the chances of failure completely depend on the membership of each parameter (for example, mobility, response time and power availability). To represent the exact degree of membership of each parameter, a fuzzy logic-based approach is undertaken. If the unpredicted failure for a Fog device is high then the Fog device will be flagged as unreliable. To handle failure for unreliable Fog devices, replication is used to ensure the robustness of the scheduling algorithm.

A predicted failure handling mechanism basically acts before the resource failure takes place. However, due to decentralised management of the Fog devices, application might fail and this is beyond our prediction. Hence, an unpredicted failures handling mechanism allows seamless application processing. Frequent unpredicted failures caused by a Fog device will trigger replication to ensure successful application execution. Thus, to ensure a reliable application processing environment, all three approaches (predicted failure, unpredicted failure and replication) need to be considered. Figure 1 shows what action will be taken after calculating the failure score.


\begin{figure}[htbp]
	\centering
	\label{figprob}
	\includegraphics[width=3in]{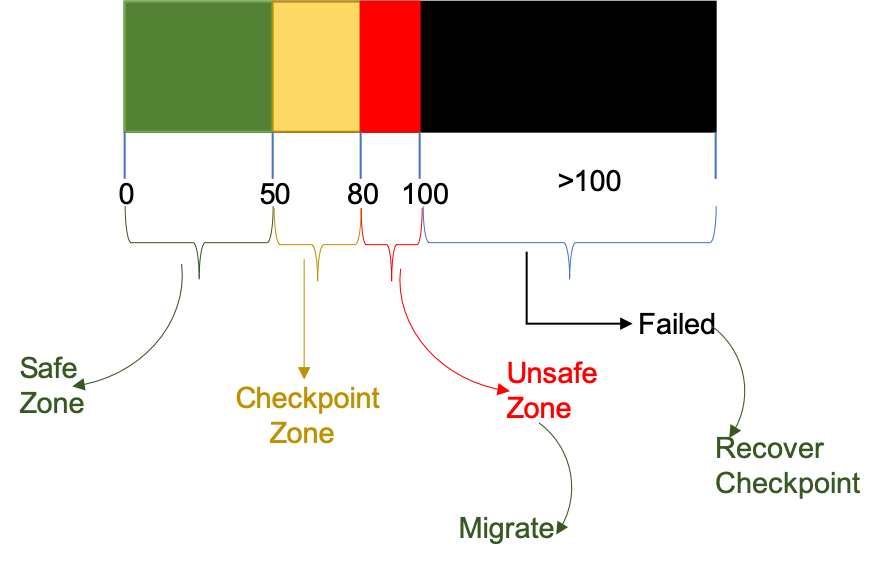}
	\caption{Proposed failure handling mechanism.}
\end{figure} 

\subsection{MRP score calculation for unpredicted failure}
To find unpredicted failure, the system always calculates the degree of failure by calculating membership of the following parameters: (i) device mobility, (ii) device response time and (iii) device power availability.


Based on the degree of failure, the system will decide how frequently checkpointing needs to be undertaken. Based on the percentage of the device movement, we can define how readily the device can be completely moved to another network. Device mobility can be represented as $D_m$ which could be 0\% to 100\%. Device response time always maps with required response time to meet the application time sensitivity. For example, to complete an application request, the device response time should be 2ms but the device is responding in 1m, therefore, the degree of failure is within the group of 0\%. On the other hand, if the device response time suddenly changed to 2.5ms then the degree of failure is within the group of 100\% since it is not meeting the application time-sensitive requirements. Device response time can be represented as $D_r$ which could be 0\% to 100\%. Similarly, the power available score can be calculated based on the power that is required to complete the submitted application. All device characteristics parameters are transformed into a normalised range [0 to 1] during fuzzification. Fuzzy logic usually includes three phases: fuzzification, fuzzy inference and defuzzification. The fuzzy sets for the above parameters are as follows:

\begin{itemize}
    \item Device mobility: $D_m \in \{ Low, Normal, High \}$
    \item Device response time: $D_r \in \{ Fast, Normal, Slow \}$
    \item Device power: $D_p \in \{ Rich, Standard, Poor \}$
\end{itemize}

Using Equation \ref{eqnorm}, the value can be normalised to fall in the interval [0 to 1].


\begin{equation}
\label{eqnorm}
    \overline{D_x}=\frac{D_x - \alpha_x}{\beta_x - \alpha_x}
\end{equation}

\begin{figure*}[t]
    \centering
    \begin{subfigure}[b]{0.32\textwidth}
        \includegraphics[width=\textwidth, height=3cm]{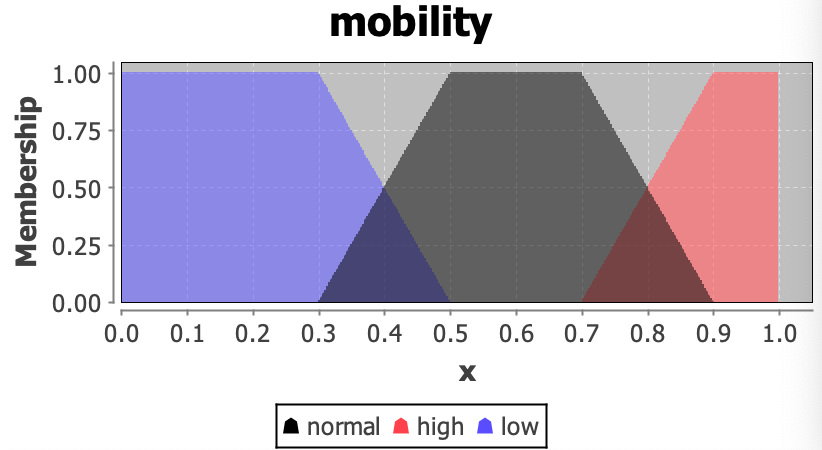}
        \caption{Mobility}
        \label{fig_ufmob}
    \end{subfigure}
    \begin{subfigure}[b]{0.32\textwidth}
        \includegraphics[width=\textwidth, height=3cm]{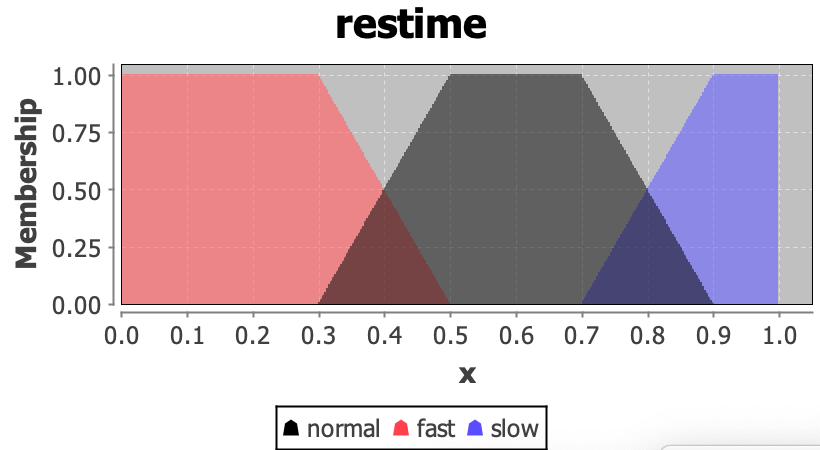}
        \caption{Response time}
        \label{fig_ufres}
    \end{subfigure}
    \begin{subfigure}[b]{0.32\textwidth}
        \includegraphics[width=\textwidth, height=3cm]{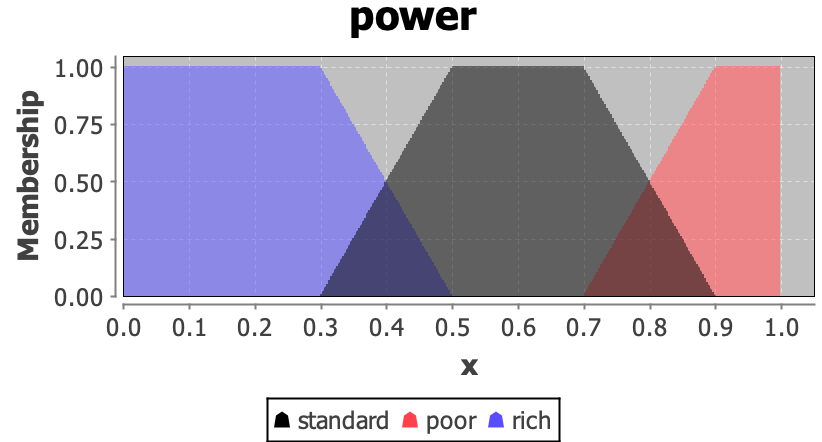}
        \caption{Power availability}
        \label{fig_ufpow}
    \end{subfigure}
  \caption{Membership of different parameters for unpredicted failure.}\label{unprepara}
\end{figure*}

\textcolor{black}{In the Equation \ref{eqnorm}, $D_x$ is the numerical value of $x$ where $x$ is either mobility, response time or power. The value of $x$ is within the range of $\alpha_x$ to $\beta_x$.} The normalised value of the parameters’ mobility, response time and power is calculated for further operation. The degree of membership of each parameter is shown in Figure \ref{unprepara}. 




\begin{figure}[htbp]
	\centering
	\label{fig_ufmrp}
	\includegraphics[width=2.5in]{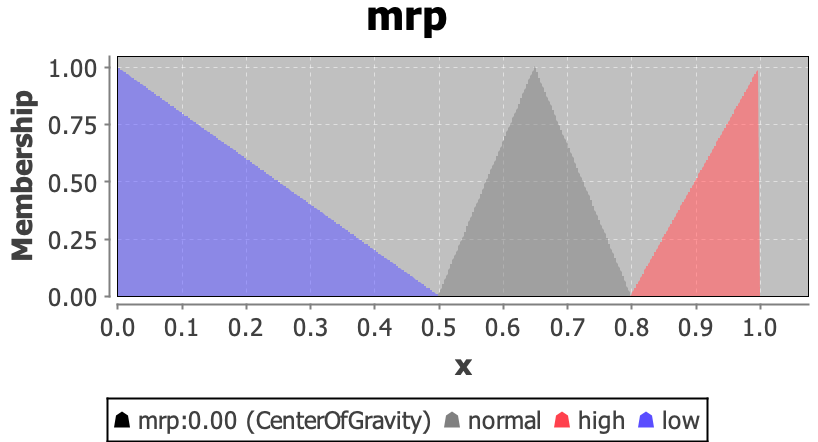}
	\caption{Membership for mrp score.}
\end{figure} 

The mobility parameter of 0\% to 50\% is considered as low mobility; 30\% to 90\% is normal mobility and 70\% to 100\% is considered to be high mobility. Until 30\% mobility membership, we consider that the system is in safe zone. However, at the point of 30\%, the mobility membership is low and is decreasing, and normal mobility membership is increasing. At the point of 50\%, the low mobility membership is 0 and normal mobility membership is 1. However, at the point of 70\% normal mobility, membership is going to decrease and will become 0 at 90\%. On the other hand, a 70\% high mobility score which starts to increase and reaches 1 at 90\% means the device is about to fail. A similar approach is employed for response time and membership of the power availability parameters. Based on the membership of each parameter, fuzzification is completed in the Fuzzy Interference System (FIS). To develop FIS we used the jfuzzylogic toolbox \cite{cingolani2013jfuzzylogic}. The membership function for low, normal and high mobility is shown in Equations \ref{eqlow}, \ref{eqnor} and \ref{eqhigh}. A similar equation is used for response time and power parameters.


\begin{equation}
\label{eqlow}
    \mu_{mL}(x)= \begin{cases} 
   0, &  x > d \\
   \frac{d-x}{d-c},   & c \leq x \leq d \\
   1,       & x < c
  \end{cases}
\end{equation}

\begin{equation}
\label{eqnor}
    \mu_{mN}(x)= \begin{cases} 
   0, &  (x < a) \ or \ (x > d) \\
   \frac{x-a}{b-a},   & a \leq x \leq b \\
   1,       & b \leq x \leq c \\
   \frac{d-x}{d-c},   & c \leq x \leq d
  \end{cases}
\end{equation}

\begin{equation}
\label{eqhigh}
    \mu_{mH}(x)= \begin{cases} 
   0, &  x < a \\
   \frac{x-a}{b-a},   & a \leq x \leq b \\
   1,       & x > b
  \end{cases}
\end{equation}

We used max function as an accumulation method by which the fuzzy outcome of a particular application is represented as $X_i$ 
\\
\textbf{Fuzzy rules:} Based on the behaviour of the system fuzzy rules have generated. If any of the parameters are high the system will not be capable of running any application. More clearly, if a system is highly mobile, there is a high chance of resource failure in that device. In the same way, if response time or power membership are high, then resource failure in that particular device will also be high. For this particular instance the rule should be as follows:

\begin{itemize}
    \item If $D_m$ is	high 	or $D_r$ is	slow	or $D_p$ is	poor	then	$UF_{{mrp}_{m}}$ is	high 
\end{itemize}

In the above rule, $UF_{{mrp}_{m}}$ represents an unpredicted failure score for application m. In order to consider some devices as being in a safe zone, all scores of all parameters should have safe zone scores which have a 0\% to 50\% variation. For this particular instance the rule is as follows:

\begin{itemize}
    \item If $D_m$ is	low 	and $D_r$ is	fast	and $D_p$ is	rich	then	$UF_{{mrp}_{m}}$ is	low
\end{itemize}

To define device membership in the checkpoint zone, mobility should be low or normal, response time should be fast or normal and power availability should be rich or standard. The mobility membership should not be high, response time should not be slow and power should not be poor, to be in the checkpointing zone. In addition, mobility should not be low, response time should not be fast and power should not be rich at the same time. To represent the situations described above, we need to define seven different rules. An example of such a rule is given as follows:

\begin{itemize}
    \item If $D_m$ is	low 	and $D_r$ is	fast	and $D_p$ is	standard 	then $UF_{{mrp}_{m}}$ is	normal 
\end{itemize}

\textbf{Fuzzy inference and defuzzification:}
To generate an mrp score we used 0\% 50\% as a low score, 50\% to 80\% is a normal score and 80\% to 100\% is a high score (See Figure 3). A Center of Gravity (CoG) defuzzification method is used for calculating the mrp score. The equation for CoG is shown in equation \ref{eqcog}. \\

\begin{equation}
\label{eqcog}
    UF_{{mrp}_{x}} = \frac{\sum_{i=1}^{n} X_i \times \mu_i}{\sum_{i=1}^{n} X_i}
\end{equation}

In the above equation n is the number of rules needing to be triggered. $\mu_i$ is the singleton value which refers to the maximum score for a particular parameter. The defuzzification value for an application is used to make decisions about application failure handling (mrp score).

\subsection{CPMNR score calculation for predicted failure}
Some failures can be predicted based on the following criteria: 
\begin{itemize}
    \item Effect on processing based on the current CPU utilisation
    \item Effect on processing based on available power in the device
    \item Effect on processing based on device mobility
    \item Effect on processing based on network communication (Device is capable of completing the request but network communication might be the cause of not meeting time-sensitive requirements)
    \item Effect on processing based on device response time
\end{itemize}

All device behaviour parameters are transformed into a normalised range [0 to 1] during fuzzification. The fuzzy sets for the above parameters are as follows:

\begin{itemize}
    \item CPU utilization: $D_{mc} \in \{ Low, Normal, High \}$
    \item Device power: $D_{mp} \in \{ Rich, Standard, Poor \}$
    \item Device mobility: $D_{mm} \in \{ Low, Normal, High \}$
    \item Network communication: $D_{mn} \in \{ Fast, Medium, Slow \}$
    \item Response time $D_{mr} \in \{ Fast, Normal, Slow \}$
\end{itemize}

Using Equation \ref{eqnor1}, the value can be normalised to fall into the interval [0 to 1].

\begin{equation}
\label{eqnor1}
    \overline{D_{mx}}=\frac{D_{mx} - \alpha_{mx}}{\beta_{mx} - \alpha_{mx}}
\end{equation}

\begin{figure*}[t]
    \centering
    \begin{subfigure}[b]{0.32\textwidth}
        \includegraphics[width=\textwidth, height=3cm]{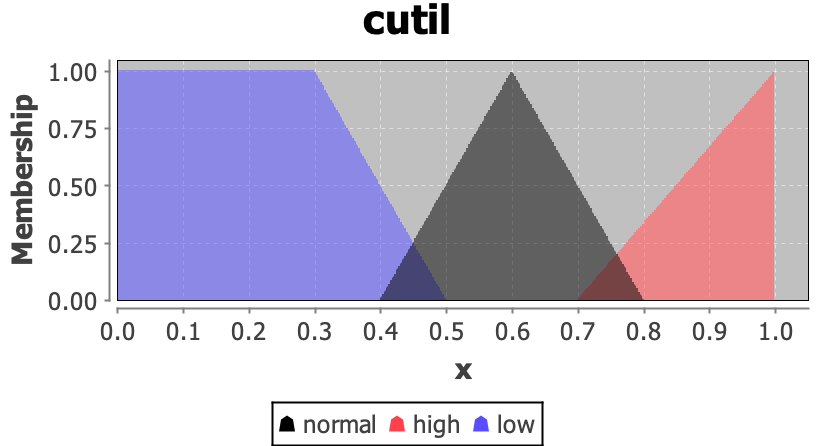}
        \caption{CPU utilisation}
        \label{fig_pcutil}
    \end{subfigure}
    \begin{subfigure}[b]{0.32\textwidth}
        \includegraphics[width=\textwidth, height=3cm]{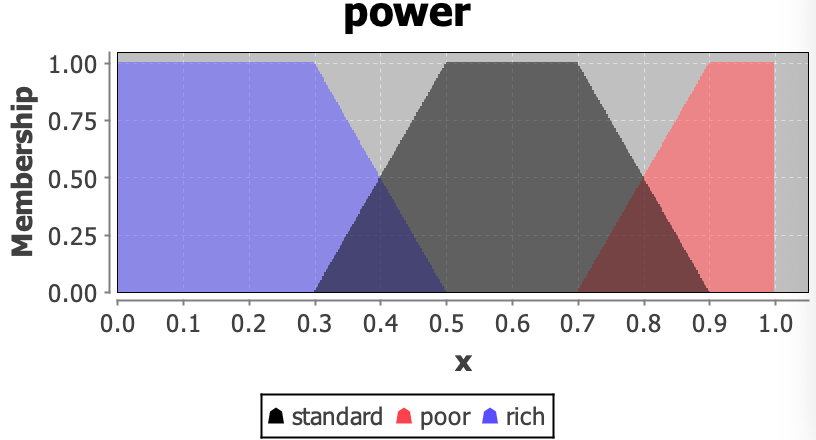}
        \caption{Power availability}
        \label{fig_pfpow}
    \end{subfigure}
    \begin{subfigure}[b]{0.32\textwidth}
        \includegraphics[width=\textwidth, height=3cm]{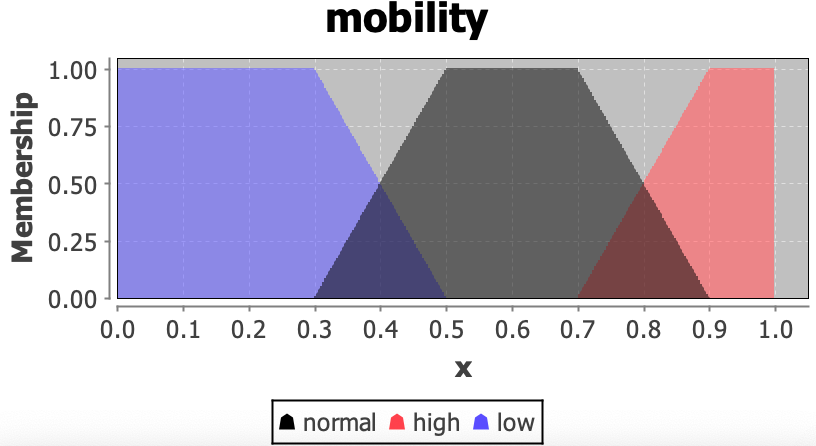}
        \caption{Mobility}
        \label{fig_pfmob}
    \end{subfigure}
     \begin{subfigure}[b]{0.32\textwidth}
        \includegraphics[width=\textwidth, height=3cm]{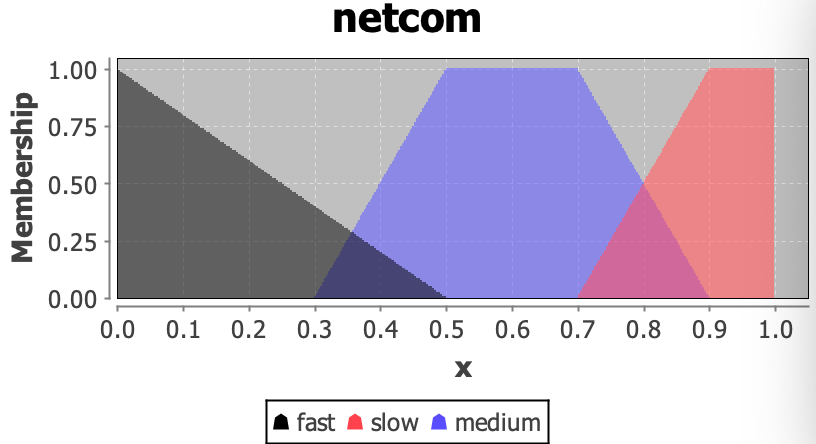}
        \caption{Network communication}
        \label{fig_pfnet}
    \end{subfigure}
    \begin{subfigure}[b]{0.32\textwidth}
        \includegraphics[width=\textwidth, height=3cm]{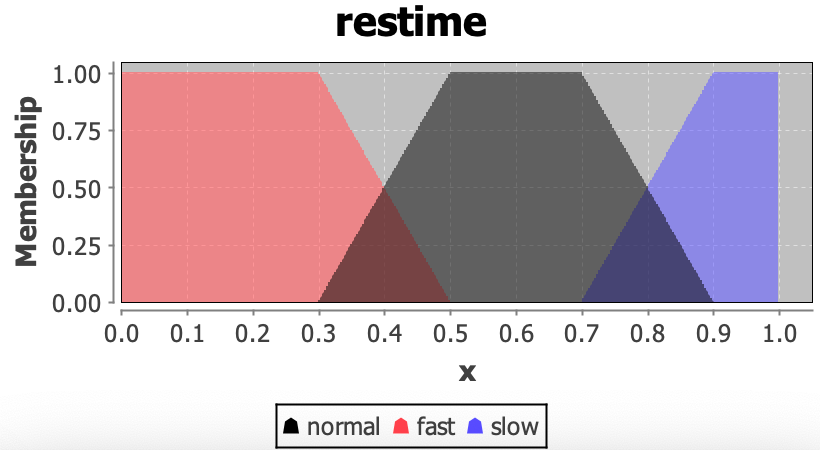}
        \caption{Response time}
        \label{fig_pfres}
    \end{subfigure}
    \begin{subfigure}[b]{0.32\textwidth}
        \includegraphics[width=\textwidth, height=3cm]{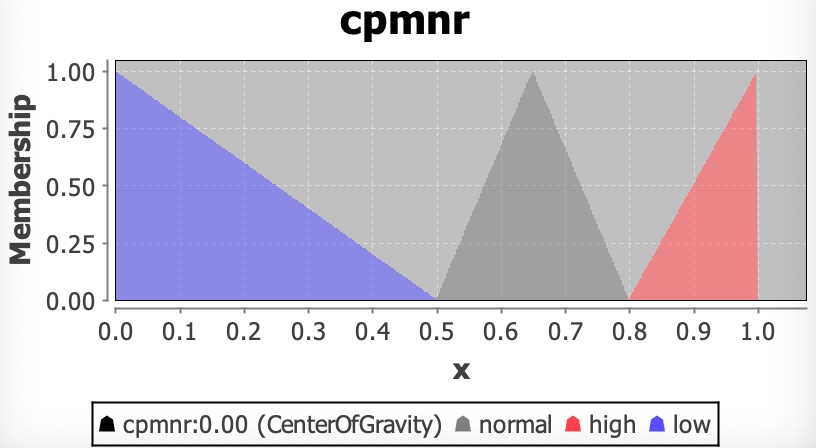}
        \caption{cpmnr score}
        \label{fig_ufcpmnr}
    \end{subfigure}
  \caption{Membership of different parameters for predicted failure and cpmnr score.}\label{prepara}
\end{figure*}

\textcolor{black}{In the Equation \ref{eqnor1}, $D_{mx}$ is the numerical value of $mx$ where $mx$ is either CPU utilisation, power, mobility, network communication or response time. The value of $mx$ is within the range of $\alpha_{mx}$ to $\beta_{mx}$.} The normalised value of  the parameters' CPU utilisation, power, mobility, network communication and response time is calculated for further operation. The degree of membership of each parameter is shown in Figures \ref{prepara}. 

The CPU utilisation parameter 0\% to 50\% is considered to be low CPU utilisation, 30\% to 90\% is normal CPU utilisation and 70\% to 100\% is considered to be high CPU utilisation. Until 30\% CPU utilisation membership, we are considering that the system is in safe zone. However, at the point of 30\% low CPU utilisation membership is decreasing and normal CPU utilisation membership is increasing. At the point of 50\% low CPU utilisation membership is 0 and normal CPU utilisation membership is 1. However, at the point of 70\%, normal CPU utilisation membership starts to decrease and will become 0 at 90\%. On the other hand, a 70\% high CPU utilisation score starting to increase and reaching 1 at 90\% means that the device is about to fail due to over utilisation of the CPU. A similar approach is employed for power availability, mobility, network communication and response time parameters. Based on the membership of each parameter, fuzzification is undertaken in the FIS system. Similar to MRP score calculation we used Equations \ref{eqlow}, \ref{eqnor} and \ref{eqhigh} for the membership function of low, normal and high for CPU utilisation, power availability, mobility, network communication and response time parameters.





Similar to the MRP score calculation we used the max function as an accumulation method by which the fuzzy outcome of a particular application is represented as  $X_i$. 
\\

\textbf{Fuzzy rules: } Based on the behaviour of the system, fuzzy rules have been generated. If any of the parameters are high the system will not be capable of running any application. More clearly, if a system has high CPU utilisation, there is a high chance of application failure in that device. In the same way, if power, mobility, network communication and response time membership are high, then application failure in that particular device will be high as well. For this particular instance the rule should be as follows:
 
\begin{itemize}
    \item If $D_{mc}$ is	high 	or $D_{mp}$ is	poor	or $D_{mm}$ is	high	or $D_{mn}$ is slow or $D_{mr}$ is slow then	$PF_{{cpmnr}_{m}}$ is	high 
\end{itemize}

In the above rule, $PF_{{cpmnr}_{m}}$ represents the unpredicted failure score for application $m$. In order to consider some devices as being in a safe zone, all scores of all parameters should have safe zone scores which are within 0\% to 50\% variation. For this particular instance the rule is as follows:

\begin{itemize}
    \item If $D_{mc}$ is	low 	and $D_{mp}$ is	rich	and $D_{mm}$ is	low	and $D_{mn}$ is fast and $D_{mr}$ is fast then	$PF_{{cpmnr}_{m}}$ is	low 
\end{itemize}

To define device membership in a checkpoint zone, CPU utilisation should be low or normal, power availability should be rich or standard, mobility should be low or normal, network communication should be fast or medium and response time should be fast or normal. The CPU utilisation membership should not be high, power should not be poor, mobility membership should not be high, network communication membership should not be slow and response time should not be slow to arrive in the checkpoint zone. Also, CPU utilisation should not be low, power should not be rich, mobility should not be low, network communication should not be fast and response time should not be fast at the same time. To represent the situations described above, we need to define 31 different rules. An example of such a rule is given as follows: 

\begin{itemize}
    \item If $D_{mc}$ is	low 	and $D_{mp}$ is	rich	and $D_{mm}$ is	low	and $D_{mn}$ is fast and $D_{mr}$ is normal then	$PF_{{cpmnr}_{m}}$ is	normal 
\end{itemize}

\textbf{Fuzzy inference and defuzzification:}
To generate the cpmnr score we used 0\% 50\% as a low score, 50\% to 80\% is a normal score and 80\% to 100\% is a high score (See Figure \ref{fig_ufcpmnr}). The Centre of Gravity (CoG) defuzzification method is used for calculating the cpmnr score. The equation for CoG is shown in Equation \ref{eqcogpf}. 

\begin{equation}
\label{eqcogpf}
    PF_{{cpmnr}_{x}} = \frac{\sum_{i=1}^{n} X_i \times \mu_i}{\sum_{i=1}^{n} X_i}
\end{equation}

In the above equation n is the number of rules needing to be triggered. $\mu_i$ is the singleton value which refers to the maximum score for a particular parameter. The defuzzification value for an application is used for making decisions about the predicted application failure handling (cpmnr score).







\subsection{Replication}
\begin{figure}[htbp]
	\centering
	\includegraphics[width=3in]{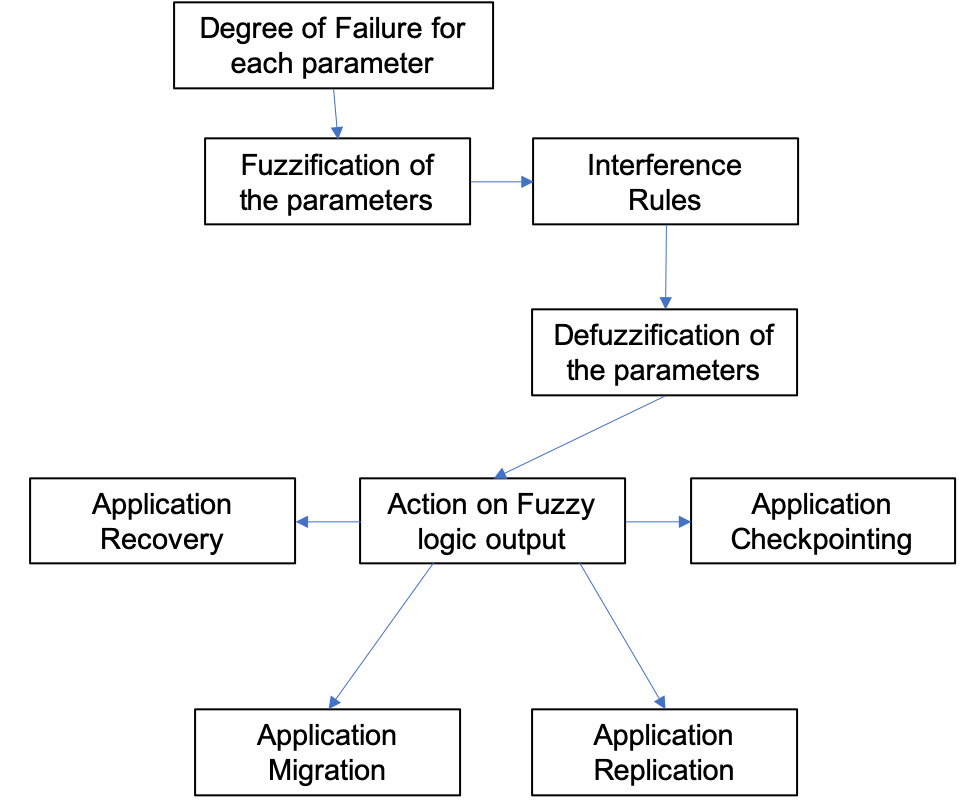}
    \caption{Failure handling process in the proposed method.}
	\label{fig_fi1}
\end{figure}

Replication of the application only applies if the rate of unpredicted (immediate) failure is high. The failure rate cannot be calculated within the execution of a few application attempts. Due to this we are considering at least 10 application executions before deciding whether replication is required or not. The overall process of the failure handling process is presented in Figure  \ref{fig_fi1}.

\begin{figure*}[htbp]
    \centering
    \begin{subfigure}[b]{0.44\textwidth}
        \includegraphics[width=\textwidth, height=5cm]{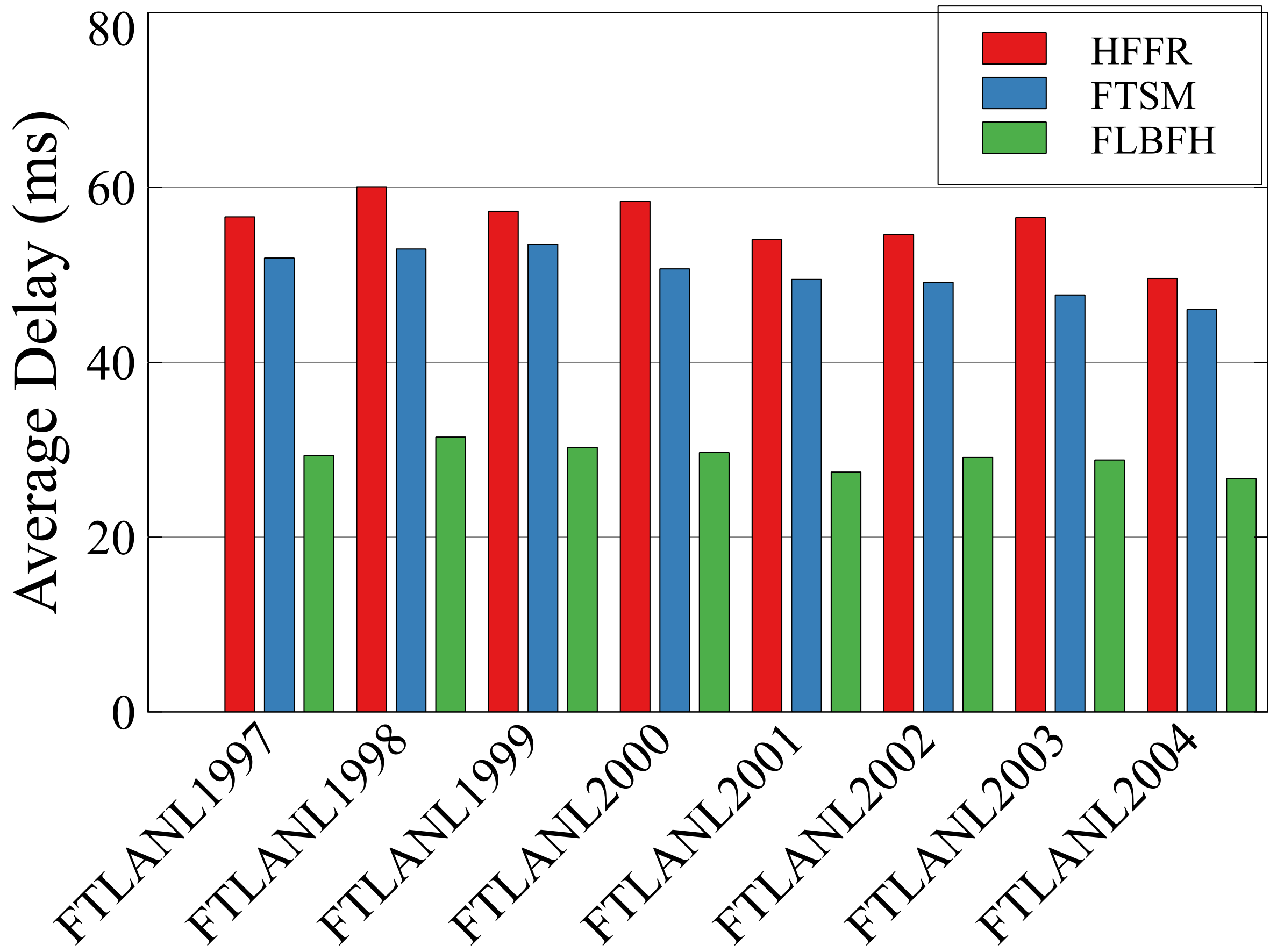}
        \caption{Average delay}
        \label{fig_01}
    \end{subfigure}
    \begin{subfigure}[b]{0.44\textwidth}
        \includegraphics[width=\textwidth, height=5cm]{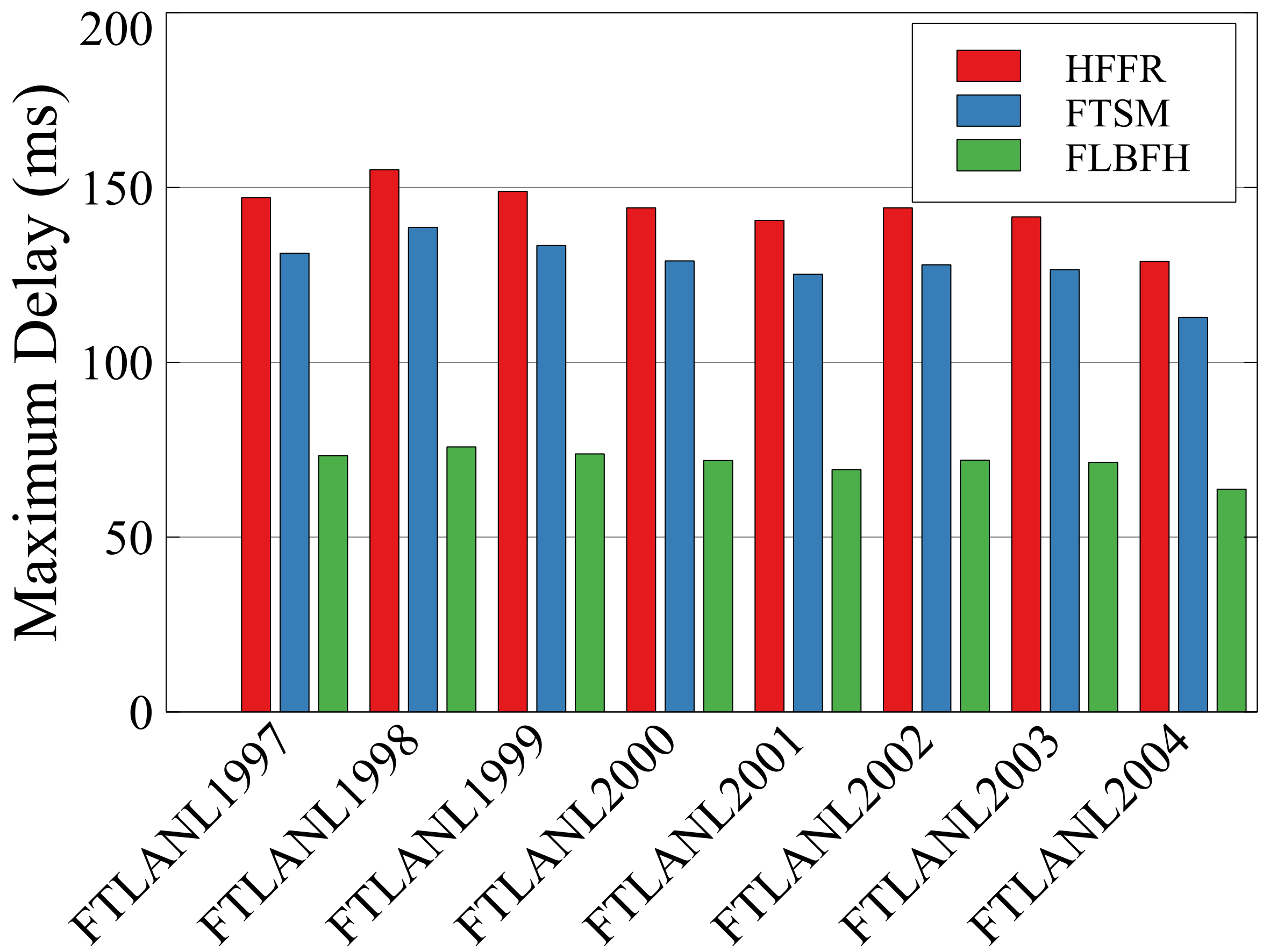}
        \caption{Maximum delay}
        \label{fig_02}
    \end{subfigure}
    \begin{subfigure}[b]{0.44\textwidth}
        \includegraphics[width=\textwidth, height=5cm]{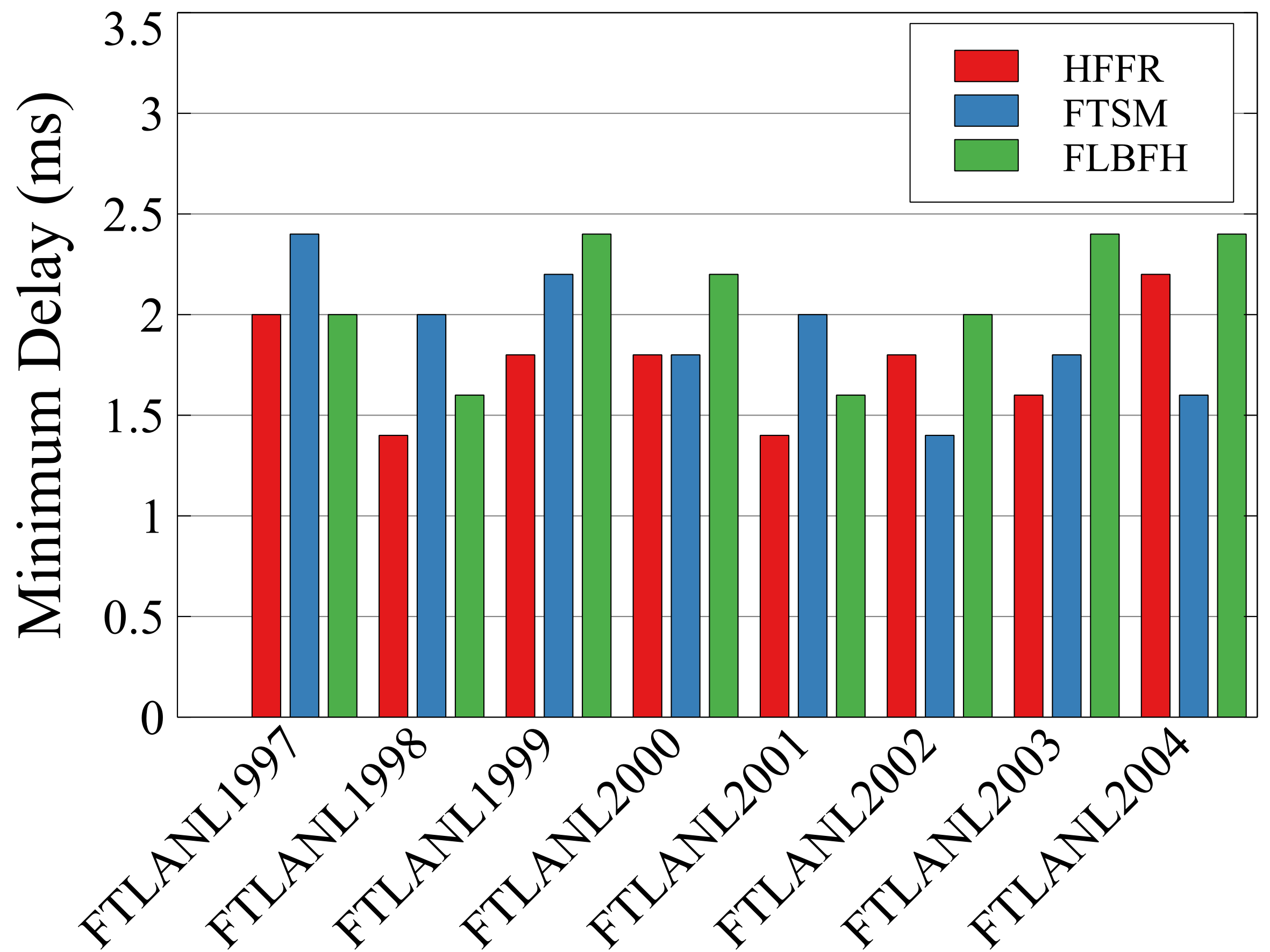}
        \caption{Minimum delay}
        \label{fig_03}
    \end{subfigure}
    \begin{subfigure}[b]{0.44\textwidth}
        \includegraphics[width=\textwidth, height=5cm]{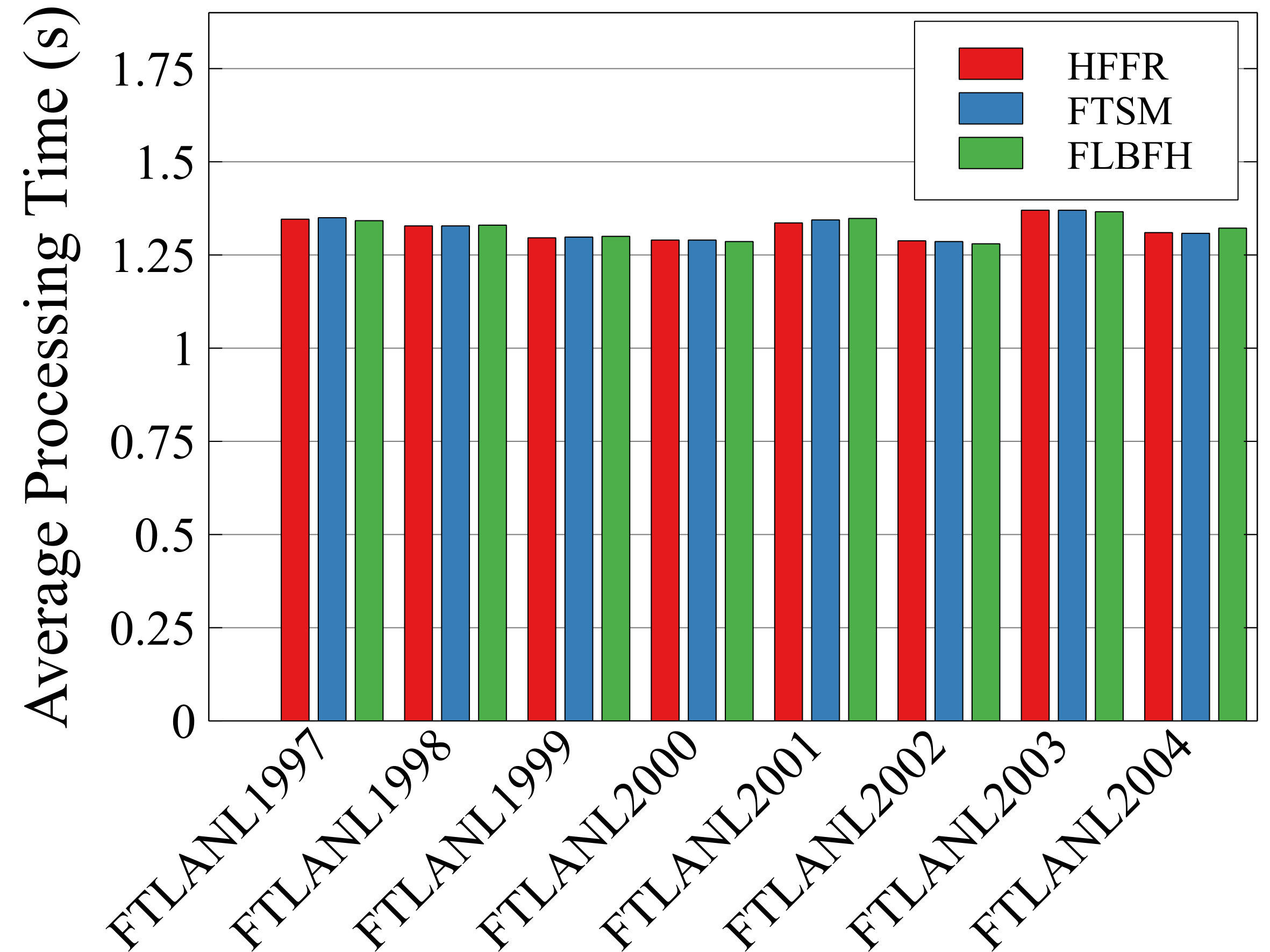}
        \caption{Average processing time}
        \label{fig_04}
    \end{subfigure}
   
  \caption{Delay and processing time for different failure traces.}\label{delapro}
\end{figure*}

\begin{algorithm}[htbp]
	\caption{Fuzzy-logic-based failure handling (FLBFH).}
	\label{raalg}
	\footnotesize
	\hspace*{\algorithmicindent} \textbf{Input: $ App[id, D_m, D_r, D_p, D_{mc}, D_{mp}, D_{mm}, D_{mn}, D_{mr}, SD_{ft}, A_c]$}  \\
	\hspace*{\algorithmicindent} \textbf{Output:$Ac_{tr}[App_{id},Actions]$} 
	\begin{algorithmic} 
		\FORALL{$App[id]$}
		\STATE Calculate degree of  changes in mobility
		\STATE Calculate degree of response time changes
		\STATE Calculate degree of power profile changes
		\STATE Calculate degree of CPU utilization changes
		\STATE Calculate degree of changes in network comm
		\STATE Calculate degree of failure ($d_f$)
		\IF {$d_f \geq 50$ \AND $d_f \leq 80$}
		\STATE $Ac_{tr}.INSERT[App_{id},Checkpoint]$
        \ELSIF {$d_f \geq 80$ \AND $d_f < 100$}
        \STATE $Ac_{tr}.INSERT[App_{id},Migrate]$
        \ELSIF {$d_f \geq 100$}
        \STATE $Ac_{tr}.INSERT[App_{id},CheckpointRecover]$
        \ENDIF
        \STATE $ASD_f$ = ($SD_{ft}$ + $d_f$)/$A_c$
        \IF {$ASD_f \geq 50$ \AND $A_c > 10$}
        \STATE $Ac_{tr}.INSERT[App_{id},Replicate]$
        \ENDIF
		\ENDFOR
	    \RETURN $Ac_{tr}[]$	
	\end{algorithmic}
\end{algorithm}

\subsection{Mapping}
If the mrp score for unpredicted failure is in the unsafe zone, then the system will migrate the application to other available Fog devices. It is obvious that the cpmnr score will also be in the unsafe zone, if the mrp score is in unsafe zone. However, the cpmnr score also considers CPU utilisation and network communication for making more accurate decisions about checkpointing and migration. If either of the two scores is in the checkpointing zone, then the application checkpointing will be triggered. However, if any of these two scores are in the failed zone, then the system will see if any replicated application is running or not. If so, the system will interact with one of the replicated applications. In the case of no running replicated application, the system will check if there any checkpoints there or not. If there are any checkpoints, then the application will recover from that checkpoint. In the case of no checkpoint and no replicated application, n, then the system will rerun the whole application which is a worse case scenario. A proposed Fuzzy-Logic-based Failure Handling (FLBFH) algorithm is presented in Algorithm  \ref{raalg}.

In Algorithm  \ref{raalg}, $SD_f$ is the score for degree of failure, $A_c$ is the app count (Total number of times task for an application is running), $SD_{ft}$ is the total score for degree of failure and $ASD_f$ is the  average score for degree of failure.









\section{Experimental Setup and Evaluation Technique}
\subsection{Failure Modelling}
\textcolor{red}{Zarza et al. fault-tolerant routing
method designed to solve a large number of dynamic permanent and non-permanent link faults}

Since no failure traces are available for the Fog, we are using failure traces from the Failure Trace Archive (FTA) \cite{javadi2013failure}. There are 27 real failure traces available in FTA. Most of those traces have two events: failed or not failed (available). Among them, only Los Alamos National Laboratory (LANL)  \cite{schroeder2009large} has failure traces with reasons such as CPU failure, power failure or network failure. Therefore, we have selected LANL failure traces to model failure in the Fog environment. LANL has failure traces for nine years (1996 to 2005) which consist of 4750 nodes that form 22 High-Performance Computing (HPC) systems \cite{schroeder2009large}. This trace has the record for every failure that takes place within the system and which needs administrator attention. We select those devices from LANL failure traces which have comparatively high failure rates compared with other devices. Those selected devices do not have failure traces for the year 1996 and 2005. Due to that, we used failure traces from 1997 to 2004.

\subsection{Experimental Setup}
For control over the environment, we chose simulation for the evaluation of the proposed method. We adopted a simulation environment and performance parameters from our previous works \cite{naha2019deadline} \cite{naha2019multi}. Also, we model a realistic Fog environment using the CloudSim \cite{calheiros2011cloudsim} toolkit, similar to our previous work.

\subsection{Evaluation Technique}
We compared proposed FLBFH with two recent works HFFR  \cite{tajiki2019software} and FTSM \cite{alarifi2019fault}. Since those two works implemented in a different simulation environment, as well as they, did not consider real failure traces, we adopted the key idea of both proposed methods to fit with our simulation environment and failure traces. We compared both methods with our proposed method in the results and discussion section to show the improvement of the FLBFH failure handling method over previously proposed methods. 


\section{Results and Discussion}
We took eight years of failure traces to perform simulations and simulate each year’s failure traces separately for HFFR, FTSM and the proposed FLBFH methods. Performance comparison of each metric is presented below in different sub-sections.

\subsection{Delay}

We have measured average, maximum and minimum delays for each task, as shown in Figure \ref{fig_01}, \ref{fig_02} and \ref{fig_03}. The average delay for the proposed FLBFH method is improved by around 52\% and 60\% for HFFR and FTSM respectively, on an average for all failure traces (Figure \ref{fig_01}). The maximum delay for the proposed FLBFH method is improved by around 50\% and 56\% for HFFR and FTSM respectively (Figure \ref{fig_02}). However, the minimum delay does not improve for the FLBFH algorithm. The minimum delay is more in FLBFH for most of the cases, as compared with HFFR and FTSM (Figure \ref{fig_03}). Since the average delay is improved for the proposed algorithm, the minimum delay will not have much effect on application processing.

\subsection{Processing time}
\begin{figure}[htbp]
	\centering
	\includegraphics[width=3in, height=2in]{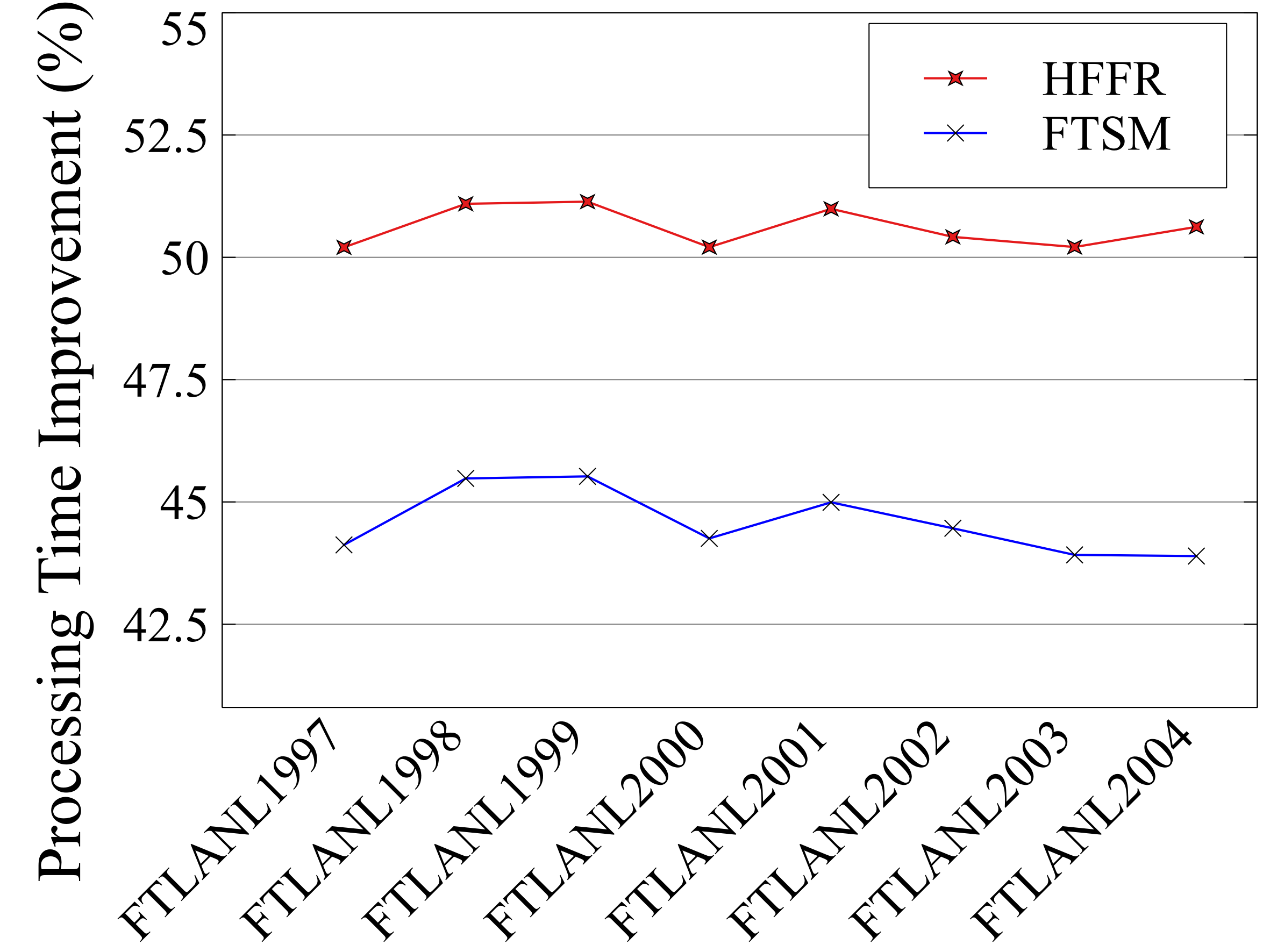}
	\caption{Improvement in processing time for different failure traces with FLBFH.}
	\label{fig_06}
\end{figure}

There is no significant difference in the average processing time for all three methods (Figure \ref{fig_04}). However, the number of failed tasks is less in the proposed FLBFH method. Since the proposed method used Fuzzy-logic based approach for failure handling and prediction, it can handle failure more efficiently, which resulting improvement in the total processing time. The total processing time improved by 51\% and 45\% for both HFFR and FTSM, compared with the FLBFH method, as shown in Figure \ref{fig_06}.

\subsection{Cost}
\begin{figure}[htbp]
	\centering
	\includegraphics[width=3in, height=2in]{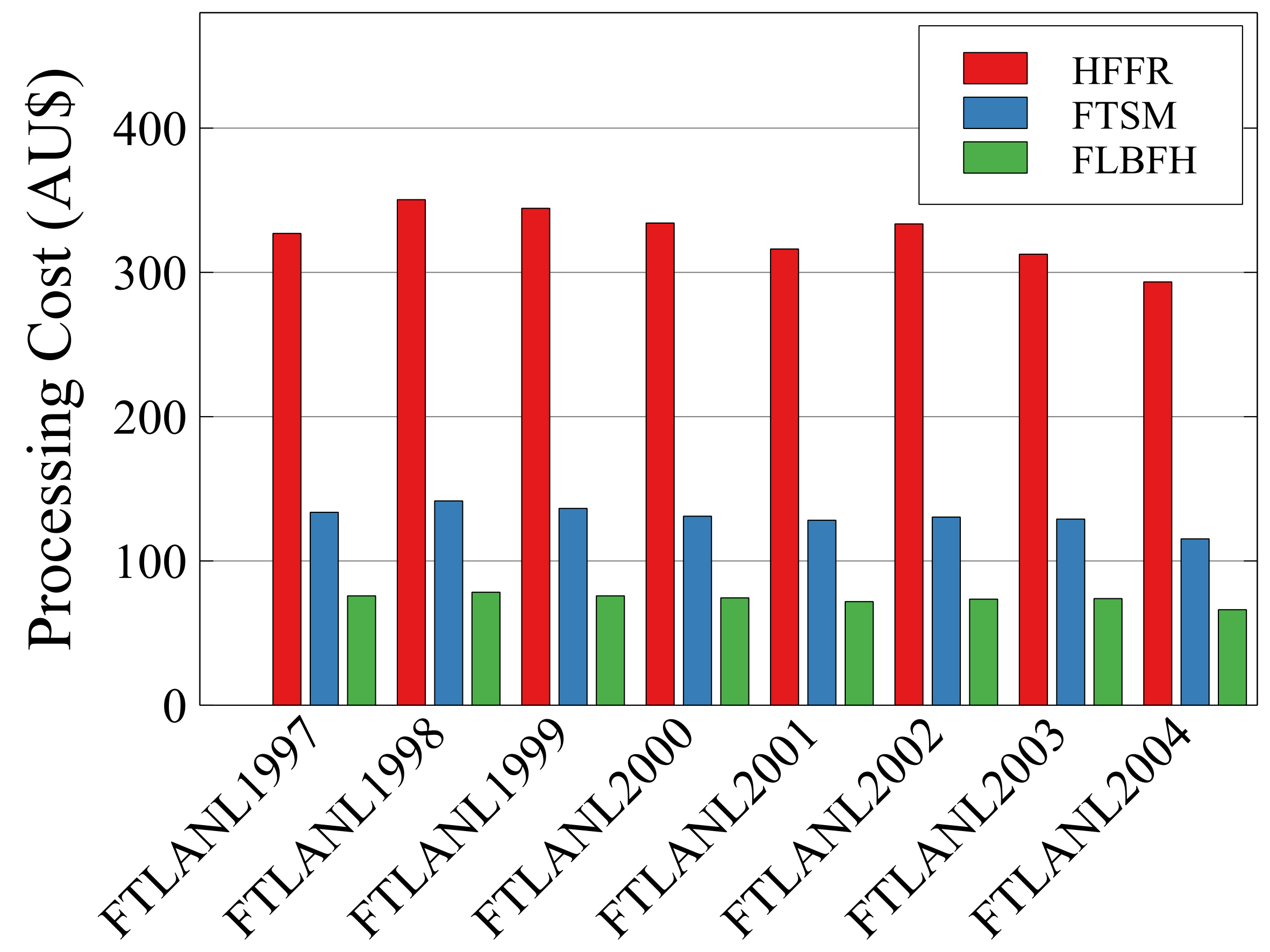}
	\caption{Processing cost for different failure traces.}
	\label{fig_05}
\end{figure}

The total processing cost is less for the proposed FLBFH method compared with HFFR and FTSM, as shown in Figure \ref{fig_05}. HFFR has around 77\% higher cost on average for all failure traces. On the other hand, FTSM has around 44\% higher cost compared with FLBFH. This indicates that the number of failed tasks is higher in HFFR and FTSM, compared with the proposed FLBFH method.

\section{Conclusion}
The Fog computing environment is highly dynamic in terms of available resources in the devices and the chances of failure are very high. This research contributes to minimising the total number of application failures due to the failure of the resources; it helps to improve delay and processing time by proposing a Fuzzy-logic-based failure handling method. The proposed failure handling method was evaluated using real failure traces from LANL. Compared with the existing failure handling approaches, we found an improvement in average delay and total processing time which are  56\% and 48\%  respectively on average. For future work, we will consider the implementation and evaluation of the proposed method in the real Fog computing environment, as well as power-aware resource allocation. The proposed method can be improved further by selecting more appropriate Fog devices which have less chance of failure. 

\bibliographystyle{IEEEtran}
\bibliography{paper}

\begin{thebibliography}{10}
\providecommand{\url}[1]{#1}
\csname url@samestyle\endcsname
\providecommand{\newblock}{\relax}
\providecommand{\bibinfo}[2]{#2}
\providecommand{\BIBentrySTDinterwordspacing}{\spaceskip=0pt\relax}
\providecommand{\BIBentryALTinterwordstretchfactor}{4}
\providecommand{\BIBentryALTinterwordspacing}{\spaceskip=\fontdimen2\font plus
\BIBentryALTinterwordstretchfactor\fontdimen3\font minus
  \fontdimen4\font\relax}
\providecommand{\BIBforeignlanguage}[2]{{%
\expandafter\ifx\csname l@#1\endcsname\relax
\typeout{** WARNING: IEEEtran.bst: No hyphenation pattern has been}%
\typeout{** loaded for the language `#1'. Using the pattern for}%
\typeout{** the default language instead.}%
\else
\language=\csname l@#1\endcsname
\fi
#2}}
\providecommand{\BIBdecl}{\relax}
\BIBdecl

\bibitem{naha2018Fog}
R.~K. Naha, S.~Garg, D.~Georgakopoulos, P.~P. Jayaraman, L.~Gao, Y.~Xiang, and
  R.~Ranjan, ``Fog computing: survey of trends, architectures, requirements,
  and research directions,'' \emph{IEEE access}, vol.~6, pp. 47\,980--48\,009,
  2018.

\bibitem{battula2019micro}
S.~K. Battula, S.~Garg, R.~K. Naha, P.~Thulasiraman, and R.~Thulasiram, ``A
  micro-level compensation-based cost model for resource allocation in a fog
  environment,'' \emph{Sensors}, vol.~19, no.~13, p. 2954, 2019.

\bibitem{yi2015survey}
S.~Yi, C.~Li, and Q.~Li, ``A survey of fog computing: concepts, applications
  and issues,'' in \emph{Proceedings of the 2015 workshop on mobile big
  data}.\hskip 1em plus 0.5em minus 0.4em\relax ACM, 2015, pp. 37--42.

\bibitem{bermbach2017research}
D.~Bermbach, F.~Pallas, D.~G. P{\'e}rez, P.~Plebani, M.~Anderson, R.~Kat, and
  S.~Tai, ``A research perspective on fog computing,'' in \emph{International
  Conference on Service-Oriented Computing}.\hskip 1em plus 0.5em minus
  0.4em\relax Springer, 2017, pp. 198--210.

\bibitem{buyya2009cloud}
R.~Buyya, C.~S. Yeo, S.~Venugopal, J.~Broberg, and I.~Brandic, ``Cloud
  computing and emerging it platforms: Vision, hype, and reality for delivering
  computing as the 5th utility,'' \emph{Future Generation computer systems},
  vol.~25, no.~6, pp. 599--616, 2009.

\bibitem{gill2018failure}
S.~S. Gill and R.~Buyya, ``Failure management for reliable cloud computing: A
  taxonomy, model and future directions,'' \emph{Computing in Science \&
  Engineering}, pp. 1--10, 2018.

\bibitem{sharma2016reliability}
Y.~Sharma, B.~Javadi, W.~Si, and D.~Sun, ``Reliability and energy efficiency in
  cloud computing systems: Survey and taxonomy,'' \emph{Journal of Network and
  Computer Applications}, vol.~74, pp. 66--85, 2016.

\bibitem{samant2004topology}
K.~Samant and S.~Bhattacharyya, ``Topology, search, and fault tolerance in
  unstructured p2p networks,'' in \emph{37th Annual Hawaii International
  Conference on System Sciences, 2004. Proceedings of the}.\hskip 1em plus
  0.5em minus 0.4em\relax IEEE, 2004, pp. 1--6.

\bibitem{vanthournout2004building}
K.~Vanthournout, G.~Deconinck, and R.~Belmans, ``Building dependable
  peer-to-peer systems,'' in \emph{DSN 2004 Workshop on Architecting Dependable
  Systems}, 2004, pp. 297--301.

\bibitem{lin2007fault}
J.-W. Lin, M.-F. Yang, and J.~Tsai, ``Fault tolerance for super-peers of p2p
  systems,'' in \emph{13th Pacific Rim International Symposium on Dependable
  Computing (PRDC 2007)}.\hskip 1em plus 0.5em minus 0.4em\relax IEEE, 2007,
  pp. 107--114.

\bibitem{mawlood2010p2p}
A.-R. Mawlood-Yunis, M.~Weiss, and N.~Santoro, ``From p2p to reliable semantic
  p2p systems,'' \emph{Peer-to-peer networking and applications}, vol.~3,
  no.~4, pp. 363--381, 2010.

\bibitem{li2006exploit}
Y.~Li and Z.~Lan, ``Exploit failure prediction for adaptive fault-tolerance in
  cluster computing,'' in \emph{Sixth IEEE International Symposium on Cluster
  Computing and the Grid (CCGRID'06)}, vol.~1.\hskip 1em plus 0.5em minus
  0.4em\relax IEEE, 2006, pp. 1--8.

\bibitem{weiss1998learning}
G.~M. Weiss and H.~Hirsh, ``Learning to predict rare events in event
  sequences,'' in \emph{KDD}, 1998, pp. 359--363.

\bibitem{vilalta2002predicting1}
R.~Vilalta and S.~Ma, ``Predicting rare events in temporal domains,'' in
  \emph{2002 IEEE International Conference on Data Mining, 2002.
  Proceedings.}\hskip 1em plus 0.5em minus 0.4em\relax IEEE, 2002, pp.
  474--481.

\bibitem{sahoo2003critical}
R.~K. Sahoo, A.~J. Oliner, I.~Rish, M.~Gupta, J.~E. Moreira, S.~Ma, R.~Vilalta,
  and A.~Sivasubramaniam, ``Critical event prediction for proactive management
  in large-scale computer clusters,'' in \emph{Proceedings of the ninth ACM
  SIGKDD international conference on Knowledge discovery and data
  mining}.\hskip 1em plus 0.5em minus 0.4em\relax ACM, 2003, pp. 426--435.

\bibitem{vilalta2002predicting}
R.~Vilalta and S.~Ma, ``Predicting rare events in temporal domains using
  associative classification rules,'' \emph{Technical Report}, pp. 426--435,
  2002.

\bibitem{hoffmann2011advanced}
G.~A. Hoffmann, F.~Salfner, and M.~Malek, \emph{Advanced Failure Prediction in
  Complex Software Systems}.\hskip 1em plus 0.5em minus 0.4em\relax
  Humboldt-Universit{\"a}t zu Berlin, Mathematisch-Naturwissenschaftliche
  Fakult{\"a}t II, Institut f{\"u}r Informatik, 2011.

\bibitem{leangsuksun2004failure}
C.~Leangsuksun, T.~Liu, T.~Rao, S.~Scott, and R.~Libby, ``A failure predictive
  and policy-based high availability strategy for linux high performance
  computing cluster,'' in \emph{The 5th LCI International Conference on Linux
  Clusters: The HPC Revolution}.\hskip 1em plus 0.5em minus 0.4em\relax
  Citeseer, 2004, pp. 18--20.

\bibitem{castelli2001proactive}
V.~Castelli, R.~E. Harper, P.~Heidelberger, S.~W. Hunter, K.~S. Trivedi,
  K.~Vaidyanathan, and W.~P. Zeggert, ``Proactive management of software
  aging,'' \emph{IBM Journal of Research and Development}, vol.~45, no.~2, pp.
  311--332, 2001.

\bibitem{oliner2005probabilistic}
A.~J. Oliner, L.~Rudolph, R.~K. Sahoo, J.~E. Moreira, and M.~Gupta,
  ``Probabilistic qos guarantees for supercomputing systems,'' in \emph{2005
  International Conference on Dependable Systems and Networks (DSN'05)}.\hskip
  1em plus 0.5em minus 0.4em\relax IEEE, 2005, pp. 634--643.

\bibitem{chakravorty2005proactive}
S.~Chakravorty, C.~Mendes, and L.~Kal{\'e}, ``Proactive fault tolerance in
  large systems,'' in \emph{HPCRI Workshop in conjunction with HPCA}, 2005, pp.
  1--7.

\bibitem{hwang2003grid}
S.~Hwang and C.~Kesselman, ``Grid workflow: a flexible failure handling
  framework for the grid,'' in \emph{High Performance Distributed Computing,
  2003. Proceedings. 12th IEEE International Symposium on}.\hskip 1em plus
  0.5em minus 0.4em\relax IEEE, 2003, pp. 126--137.

\bibitem{jin2003fault}
H.~Jin, D.~Zou, H.~Chen, J.~Sun, and S.~Wu, ``Fault-tolerant grid architecture
  and practice,'' \emph{Journal of Computer Science and Technology}, vol.~18,
  no.~4, p. 423, 2003.

\bibitem{lee2005resource}
H.~Lee, K.~Chung, S.~Chin, J.~Lee, D.~Lee, S.~Park, and H.~Yu, ``A resource
  management and fault tolerance services in grid computing,'' \emph{Journal of
  Parallel and Distributed Computing}, vol.~65, no.~11, pp. 1305--1317, 2005.

\bibitem{kandaswamy2008fault}
G.~Kandaswamy, A.~Mandal, and D.~A. Reed, ``Fault tolerance and recovery of
  scientific workflows on computational grids,'' in \emph{2008 Eighth IEEE
  International Symposium on Cluster Computing and the Grid (CCGRID)}.\hskip
  1em plus 0.5em minus 0.4em\relax IEEE, 2008, pp. 777--782.

\bibitem{khoo2010pro}
B.~B. Khoo and B.~Veeravalli, ``Pro-active failure handling mechanisms for
  scheduling in grid computing environments,'' \emph{Journal of Parallel and
  Distributed Computing}, vol.~70, no.~3, pp. 189--200, 2010.

\bibitem{battula2019efficient}
S.~K. Battula, S.~Garg, J.~Montgomery, and B.~H. Kang, ``An efficient resource
  monitoring service for fog computing environments,'' \emph{IEEE Transactions
  on Services Computing}, 2019.

\bibitem{sharma2019failure}
Y.~Sharma, W.~Si, D.~Sun, and B.~Javadi, ``Failure-aware energy-efficient vm
  consolidation in cloud computing systems,'' \emph{Future Generation Computer
  Systems}, vol.~94, pp. 620--633, 2019.

\bibitem{luo2019improving}
L.~Luo, S.~Meng, X.~Qiu, and Y.~Dai, ``Improving failure tolerance in
  large-scale cloud computing systems,'' \emph{IEEE Transactions on
  Reliability}, vol.~68, no.~2, pp. 620--632, 2019.

\bibitem{buyya2019manifesto}
R.~Buyya, S.~N. Srirama, G.~Casale, R.~Calheiros, Y.~Simmhan, B.~Varghese,
  E.~Gelenbe, B.~Javadi, L.~M. Vaquero, M.~A. Netto \emph{et~al.}, ``A
  manifesto for future generation cloud computing: research directions for the
  next decade,'' \emph{ACM computing surveys (CSUR)}, vol.~51, no.~5, p. 105,
  2019.

\bibitem{liu2017framework}
Y.~Liu, J.~E. Fieldsend, and G.~Min, ``A framework of fog computing:
  Architecture, challenges, and optimization,'' \emph{IEEE Access}, vol.~5, pp.
  25\,445--25\,454, 2017.

\bibitem{alarifi2019fault}
A.~Alarifi, F.~Abdelsamie, and M.~Amoon, ``A fault-tolerant aware scheduling
  method for fog-cloud environments,'' \emph{PLOS ONE}, vol.~14, no.~10, pp.
  1--24, 2019.

\bibitem{tajiki2019software}
M.~M. Tajiki, M.~Shojafar, B.~Akbari, S.~Salsano, and M.~Conti, ``Software
  defined service function chaining with failure consideration for fog
  computing,'' \emph{Concurrency and Computation: Practice and Experience},
  vol.~31, no.~8, pp. 1--14, 2019.

\bibitem{kai2016Fog}
K.~Kai, W.~Cong, and L.~Tao, ``Fog computing for vehicular ad-hoc networks:
  paradigms, scenarios, and issues,'' \emph{the journal of China Universities
  of Posts and Telecommunications}, vol.~23, no.~2, pp. 56--96, 2016.

\bibitem{madsen2013reliability}
H.~Madsen, B.~Burtschy, G.~Albeanu, and F.~Popentiu-Vladicescu, ``Reliability
  in the utility computing era: Towards reliable fog computing,'' in \emph{2013
  20th International Conference on Systems, Signals and Image Processing
  (IWSSIP)}.\hskip 1em plus 0.5em minus 0.4em\relax IEEE, 2013, pp. 43--46.

\bibitem{de2018fault}
J.~P. de~Araujo~Neto, D.~M. Pianto, and C.~G. Ralha, ``A fault-tolerant
  agent-based architecture for transient servers in fog computing,'' in
  \emph{2018 30th International Symposium on Computer Architecture and High
  Performance Computing (SBAC-PAD)}.\hskip 1em plus 0.5em minus 0.4em\relax
  IEEE, 2018, pp. 282--289.

\bibitem{puliafito2018companion}
C.~Puliafito, E.~Mingozzi, C.~Vallati, F.~Longo, and G.~Merlino, ``Companion
  fog computing: Supporting things mobility through container migration at the
  edge,'' in \emph{2018 IEEE International Conference on Smart Computing
  (SMARTCOMP)}.\hskip 1em plus 0.5em minus 0.4em\relax IEEE, 2018, pp. 97--105.

\bibitem{cingolani2013jfuzzylogic}
P.~Cingolani and J.~Alcal{\'a}-Fdez, ``jfuzzylogic: a java library to design
  fuzzy logic controllers according to the standard for fuzzy control
  programming,'' \emph{International Journal of Computational Intelligence
  Systems}, vol.~6, no. sup1, pp. 61--75, 2013.

\bibitem{javadi2013failure}
B.~Javadi, D.~Kondo, A.~Iosup, and D.~Epema, ``The failure trace archive:
  Enabling the comparison of failure measurements and models of distributed
  systems,'' \emph{Journal of Parallel and Distributed Computing}, vol.~73,
  no.~8, pp. 1208--1223, 2013.

\bibitem{schroeder2009large}
B.~Schroeder and G.~Gibson, ``A large-scale study of failures in
  high-performance computing systems,'' \emph{IEEE transactions on Dependable
  and Secure Computing}, vol.~7, no.~4, pp. 337--350, 2009.

\bibitem{naha2019deadline}
R.~K. Naha, S.~Garg, A.~Chan, and S.~K. Battula, ``Deadline-based dynamic
  resource allocation and provisioning algorithms in fog-cloud environment,''
  \emph{Future Generation Computer Systems}, vol. 104, pp. 131--141, 2020.

\bibitem{naha2019multi}
R.~K. Naha and S.~Garg, ``Multi-criteria-based dynamic user behaviour aware
  resource allocation in fog computing,'' \emph{arXiv preprint}, 2019.

\bibitem{calheiros2011cloudsim}
R.~N. Calheiros, R.~Ranjan, A.~Beloglazov, C.~A. De~Rose, and R.~Buyya,
  ``Cloudsim: a toolkit for modeling and simulation of cloud computing
  environments and evaluation of resource provisioning algorithms,''
  \emph{Software: Practice and experience}, vol.~41, no.~1, pp. 23--50, 2011.

\end{thebibliography}

\end{document}